\documentclass{llncs}
\usepackage{llncsdoc}
\usepackage{graphicx}
\usepackage{bm}
\usepackage{graphicx}

\begin{document}

\title{Temporal Analysis of Influence to Predict Users' Adoption in Online Social Networks}

\author{Ericsson Marin\inst{1} \and Ruocheng Guo\inst{1} \and Paulo Shakarian\inst{1}}
\institute{Arizona State University, Tempe, Arizona, USA}

\maketitle

\begin{abstract}
Different measures have been proposed to predict whether individuals will adopt a new behavior in online social networks, given the influence produced by their neighbors. In this paper, we show one can achieve significant improvement over these standard measures, extending them to consider a pair of time constraints. These constraints provide a better proxy for social influence, showing a stronger correlation to the probability of influence as well as the ability to predict influence.
\end{abstract}

\vspace{-10pt}
\section{Introduction}
\begin{figure}[!h]
\centering
\vspace{-15pt}
\includegraphics[scale=0.4,keepaspectratio]{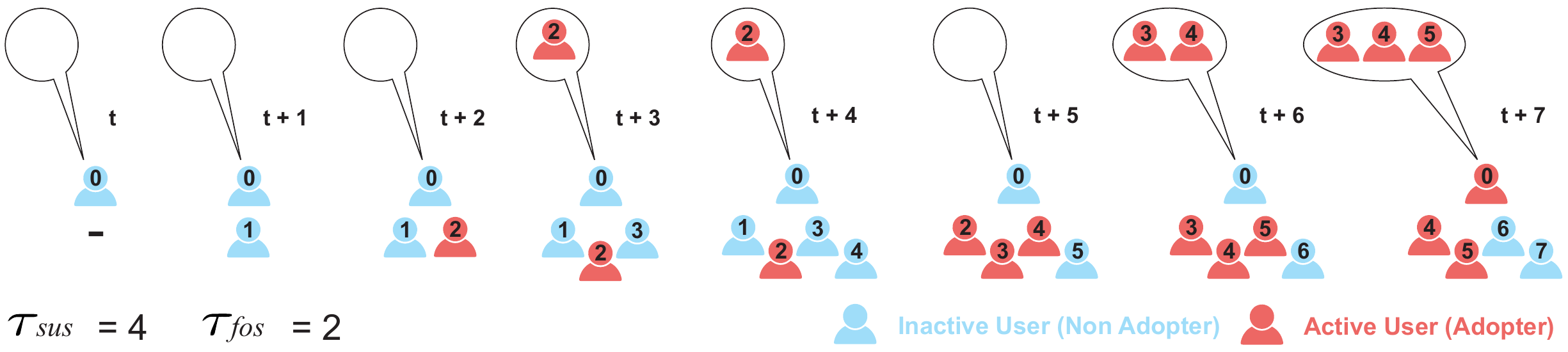}
 \vspace{-7pt}
 \caption{At time $t$, the ego node 0 has no neighbors. At $t+1$, he has 1 neighbor (node 1 at the bottom). From this moment, node 0 will be aware about node 1' actions. At $t+2$, node 0 has 2 neighbors, nodes 1, 2. This cumulative process continues until $t+5$ when node 1 is no more a neighbor of node 0, since $\tau_{sus}$ was defined as 4. After this time limit, node 0 cannot visualize actions of node 1 anymore. The illustration also shows the node 0's memory inside the balloons. As we made $\tau_{fos}=2$, node 2 (activated at $t+2$) fades away from node 0's memory after $t+4$, when node 0 is no longer influenced by him. Therefore, at $t+7$, node 0 is activated only by nodes 3, 4 and 5.}
 \vspace{-6pt}
\label{fig:Timeline}
\end{figure}

\vspace{-10pt}
Research has shown that measures which leverage the people's ego network correlate with influence - the confidence at which their neighbors adopt a new behavior \cite{Zhang:2013}. In this paper, we introduce two time constraints to improve these measures: \textit{Susceptible Span} and \textit{Forgettable Span}.  \textit{Susceptible Span} ($\tau_{sus}$) refers to the interval when people receive social signals from their neighbors (possible influencing actions), blinding individuals to no more interesting connections.  \textit{Forgettable Span} ($\tau_{fos}$) refers to the interval before an influencer's action is forgotten by his neighbors, due to human brain limitation. These constraints define evolving graphs where influence is better measured, as illustrated in Figure \ref{fig:Timeline}.

\vspace{-12pt}


The contributions of this paper are: we introduce a framework to consider $\tau_{sus}$ and $\tau_{fos}$ in social influence;  we examine the correlation of 10 social network measures to influence under different conditions;  we compare the adoption prediction performance of our method with others \cite{Zhang:2013,Fink:2016,Goyal:2010}, showing relevant improvements. For instance, we obtained up to 92.31\% gain in correlation of a simple count of the ``active'' neighbors with the probability of influence.  Considering adoption prediction, F1 score improves from 0.606 (using the state-of-the-art \cite{Zhang:2013}) to 0.689 for active neighbors. Similar results are found for the other measures analysed.

This paper is structured as follows: Section II presents the related work. Section III formalizes a framework to consider the time constraints in our networks. Section IV presents the experimental setup to produce samples. Section V introduces the social influence measures and their corresponding gains in correlation coefficient. Section VI details the classification experiments and results for the adoption prediction problem. Finally, section VII concludes the paper. 
\vspace{-10pt}
\section{Related Work}
\vspace{-8pt}
Many works have been proposed to measure social influence and predict users' adoption. For instance, the seminal work of Kempe et al. \cite{Kempe:2003} describes two popular models for diffusion in social networks that were generalized to the General Threshold Model. In this model, the collective influence from a node's infected neighbors will trigger his infection once his threshold is exceeded. Later, Goyal et al. \cite{Goyal:2010} leveraged a variety of models based on pair-wise influence probability, finding the probability of adoption increases with more adopters amongst friends.  With an alternative approach, Zhang et al. \cite{Zhang:2013} proposed the \textit{influence locality}, developing two instantiated functions based on pair-wise influence but also on structural diversity to predict adoptions.  Comparing two different perspectives, Fink et al. \cite{Fink:2016} proposed probabilistic contagion models for simple and complex contagion, with the later producing a superior fit for themed hashtags.

In these works, the authors slightly explored the dynamic aspect of social influence. Here, we take the next steps to apply a pair of time constraints to our networks, finding that influence is better measured and predicted dynamically.
\vspace{-10pt}
\section{Framework for Consideration of Time Constraints}
\vspace{-8pt}
In this section, we describe the notations of this work. We denote a set of users $V$, as the nodes in a directed network $G=(V,E)$, a set of topics (hashtags) $\Theta$, and a set of discrete time points $T$. We will use the symbols $v,\theta,t$ to represent a specific node, topic and time point.  With nodes being active or inactive w.r.t $\theta$, an active node (adopter) is a user who retweeted a tweet with $\theta$.  We denote an activity log $\mathcal{A}$ (containing all retweets) as a set of tuples of the form $\langle v_1, v_2, \theta, t\rangle$, where $v_1, v_2 \in V$. It describes that ``$v_1$ adopted $\theta$ retweeting $v_2$ at time $t$'', creating a directed edge $(v_1,v_2) \in E$. The intuition behind this edge is that $v_1$ can be influenced by $v_2$ with respect to $\theta'$, if $v_2$ eventually adopts $\theta'$ after $t$.

Finally, we integrate into our model the two proposed time constraints $\tau_{sus}$ and $\tau_{fos}$.  Due to them, the neighborhood of a user can change over time, affecting the social influence measures that result in his decision to adopt a topic.  This way, we define the set of neighbors of a node $v$ at time $t$ as:

\vspace{-16.5pt}
\begin{eqnarray*}
\eta_{v,t} &= \{ v' | & \exists \langle v,v',\theta, t' \rangle \in \mathcal{A} \textit{, s.t. } t' \leq t \textit{ and } t-t' \leq \tau_{sus} \}
\end{eqnarray*}
$\eta_{v,t}$ is the set of users whose adoptions since $t-\tau_{sus}$ until $t$ will be presented to $v$. After $t'+\tau_{sus}$, the adoptions of $v'$ will not influence $v$. Then, we introduce $\eta_{v,t}^{\theta}$ as the set of users that can influence $v$ to adopt $\theta$ at time $t$ as:
\vspace{-5.5pt}
\fontsize{8}{12}\selectfont
\begin{eqnarray*}
\eta_{v,t}^{\theta} &= \{ v' \in \eta_{v,t'} | & \exists \langle v',v'',\theta, t'' \rangle \in \mathcal{A} \textit{, s.t. }  t' \leq t'' \textit{, } t''- t' \leq \tau_{sus} \textit{, } t'' \leq t \textit{ and } t-t'' \leq \tau_{fos} \}
\end{eqnarray*}
\normalsize

Consequently, after $t''+\tau_{fos}$, the fact that $v'$ adopted $\theta$ is forgotten by $v$, with $v'$ no more influencing $v$ in terms of $\theta$. Using these generated dynamic networks, we want to measure the influence produced by the individuals' active neighbors.

\vspace{-12pt}
\section{Experimental Setup}
\label{sec:setup}
\vspace{-11pt}
This section details our dataset, how we collect samples using different values for the time constraints, which filters of users' activity are applied, and how we measure correlation of our features with probability of adoption.

\vspace{3pt}

\noindent\textbf{Dataset description}. The dataset we use is provided by \cite{Weng:2013}. It contains 1,687,700 retweets $(k)$, made by 314,756 users (the histogram fits a power-law with $p_k \approx k^{-1.8}$), about 226,488 hashtags on Twitter, from March 24 to April 25, 2012.

\vspace{3pt}

\noindent\textbf{Sampling}. Following previous works \cite{Zhang:2013}, we create balanced sets of samples for our experiments. For a given activity $\langle v,v'',\theta,t \rangle$ which corresponds to a positive sample, we create a negative sample uniformly getting a user $v'$ from the set: $\left\{v'| v \in \eta^{\theta}_{v',t} \wedge \langle v',u,\theta,t' \rangle \not \in \mathcal{A} , \forall u \in \eta^{\theta}_{v',t}\right\}$.  This set includes all users under influence of $v$ w.r.t $\theta$ at $t$, who did not adopted $\theta$ in our dataset.  Then, we create $ \langle v',v,\theta,t \rangle $ as the related negative sample for $ \langle v,v'',\theta,t \rangle $, keeping the same timestamp for both users to have similar intervals to accumulate influence.

\vspace{3pt}

\noindent\textbf{Filters}. In addition, we apply 4 filters to exclude users with less actions than a given threshold, as their behaviors are hardly explainable by influence measurements \cite{Fink:2016}. We label \textit{R30}, \textit{R60} for users who retweeted at least 30 or 60 times, and \textit{H20}, \textit{H40} for users who retweeted at least 20 or 40 hashtags respectively. This also enables us to test the robustness of $\tau_{sus}, \tau_{fos}$ under a variety of conditions.

\vspace{3pt}

\noindent\textbf{Correlation between measures and adoption probability}. Here, we study how each time constraint correlates with probability of adoption using the Pearson correlation coefficient.  The idea is to identify the values for $\tau_{sus}$ and $\tau_{fos} \in \left\{8, 16, 24, 48, 72, 96, 120, 144, 168, 336, 504, 720\right\}(hours)$, that produce high quality influence measurements (high positive correlation with adoption probability).  

\vspace{-13pt}

\section{Influence Measures}
\vspace{-10pt}
Table \ref{tab:features} describes the 10 measures (in 7 categories), which we use to estimate the influence in users' active neighborhood.  We define the measures based on the activity $a = \langle v,v',\theta,t \rangle$ by which we create samples.  Then, we show the gain (or loss) of correlation coefficient by heat maps, plotting the 144 combinations of $\tau_{sus}$ and $\tau_{fos}$. Cells in the right lower corner have values = 0, as $(\tau_{sus},\tau_{fos}) = (720,720)$ equals to applying no time constraints (data comprises of 720 hours).

\begin{figure}[!h]
\vspace{-21pt}
\centering
\includegraphics[scale=0.28,keepaspectratio]{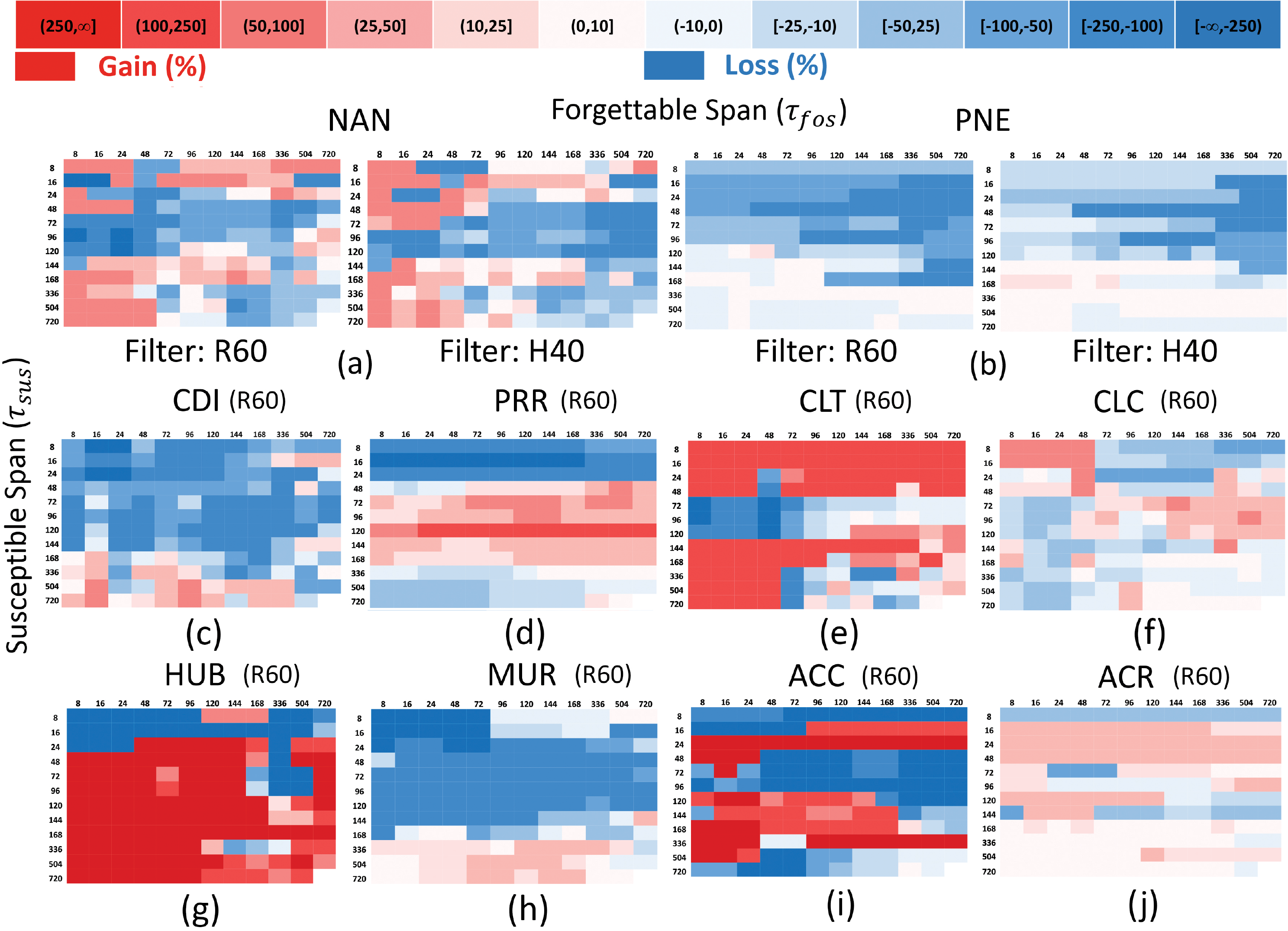}
 \vspace{-10pt}
 \caption{Gains (red) or losses (blue) of correlation between each social influence measure and adoption probability when the time constraints are applied.}
\vspace{-8pt}
\label{fig:Correlation_Gain}
\vspace{-8pt}
\end{figure}

Figure \ref{fig:Correlation_Gain}(a) shows the heat maps for filters $R60$ and $H40$ with the gain (or loss) of correlation coefficients between \textit{Number of Active Neighbors (NAN)} and probability of adoption.  Previous work with no time constraints \cite{Goyal:2010,Fink:2016} argue that a positive correlation is expected here.  Even so, many cells present gains for both filters, where combinations of $\tau_{sus}$ and $\tau_{fos}$ boost \textit{NAN}'s ability to explain users' behaviors under influence.  Moreover, hot red cells dominate the left lower region of both heat maps where $\tau_{sus}$ and $\tau_{fos}$ are relatively high and low respectively, especially when $\tau_{sus} \ge 168$ and $\tau_{fos} \le 48$, with gains in [9.84\%, 92.31\%].  Figure \ref{fig:Correlation_Gain}(b) presents the heat maps for Personal Network Exposure (PNE). Similar to NAN, hot red cells are mainly distributed where $\tau_{sus} >= 168$ and $\tau_{fos} <= 24$. Although the gains are smaller, in [1.18\%, 11.76\%], we show how PNE obtains high gains in classification performance. From this moment on, we plot the heat maps only for filter \textit{R60}, since we get similar results for $H40$.
\begin{table*}[!ht]
\centering
\tiny{
\caption{Categories of Features.}
\vspace{-2pt}
\begin{tabular}{|l|l|l|}
\hline  Category             & Feature                       & Formula \\
\hline  Connectivity         & Number of Active Neighbors \cite{Goyal:2010}    & $NAN^{\theta}_{v,t} = |\eta_{v,t}^{\theta}|$                                                                                                                  \\ \cline{2-3}
                             & Personal Network Exposure \cite{Valente:1995}   & $PNE^{\theta}_{v,t} = \frac{|\eta_{v,t}^{\theta}|}{|\eta_{v,t}|}$                                                                                             \\
\hline  Temporal             & Continuous Decay of Influence \cite{Goyal:2010} & $CDI^{\theta}_{v,t} = \sum_{u \in \eta^{\theta}_{v,t}}\mathrm{exp}(\frac{-(t_l-t_u)}{\sigma})$                                                                \\ 
                             &                               & where $t_l$ is the time when the latest neighbor in $\eta^{\theta}_{v,t}$ adopted $\theta$,                                                                                     \\ 
                             &                               & and $\sigma$ is the globally longest identified time-delay for adoption.                                                                                                        \\
\hline  Recorrence           & Previous Reposts \cite{Zafarani:2014} & $PRR^{\theta}_{v,t} = \sum_{\theta'}\sum_{u \in \eta^{\theta}_{v,t}}\sum_{t' \le t}| \langle v,u,\theta',t' \rangle |$                                                  \\ 
\hline  Transitivity         & Closed Triads \cite{Zafarani:2014}    & $CLT^{\theta}_{v,t} = \sum_{\{u,z\} \in \eta^{\theta}_{v,t}, u \neq z}f((u,z)^{\theta}_{t})$ \,\,\,\, and                                                               \\ 
                             &                               & $f((u,z)^{\theta}_{t}) = \left\{\begin{array}{l}1, \,\, if \,\, \langle u,z,\theta, t' \rangle \in \mathcal{A} \wedge t' \leq t \\0, \,\, otherwise \\ \end{array}\right.$      \\ \cline{2-3}
                             & Clustering Coefficient \cite{Zafarani:2014} & $CLC^{\theta}_{v,t} = \sum_{\{u,z\} \in \eta^{\theta}_{v,t}, u \neq z}\frac{g((u,z)^{\theta}_{t})}{|\eta_{v,t}^{\theta}|^2}$ \,\,\, and                           \\ 
                             &                               & $g((u,z)^{\theta}_{t}) = \left\{\begin{array}{l}1, \,\, if \,\, \langle u,z,\theta, t_z \rangle \in \mathcal{A} \wedge t_z \leq t \\0, \,\, otherwise \\ \end{array}\right.$   \\		
\hline  Centrality           & Hubs \cite{Zafarani:2014}     & $HUB^{\theta}_{v,t} = \sum_{u \in \eta^{\theta}_{v,t}}h(u,t)$ \,\,\,\,\,\,\,\,\,\,\,\,\,\,\,\,\,\,\,\,\,\,\,\,\,\,\,\,\,\,\,\,\,\,\,\,\, and                                    \\  
                             &                               & $h(u,t) = \left\{\begin{array}{l}1, \,\, if \sum_{\theta'}\sum_{x \in V}\sum_{t' \le t}|\langle x,u,\theta',t' \rangle| >= \gamma \\ 0, \,\, otherwise \\ \end{array}\right.$  \\
                             &                               & where $\gamma$ being the minimal number of messages retweeted. Upon                                                                                                             \\
                             &                               & some analysis, we made $\gamma$ = 104, corresponding to 0.042\% of all                                                                                                          \\
                             &                               & retweets. To reach this value, users should be retweeted at least.                                                                                                              \\
\hline  Reciprocity          & Mutual Reposts \cite{Zafarani:2014} & $MUR^{\theta}_{v,t} = \sum_{u \in \eta^{\theta}_{v,t}}i(u,t)$  \,\,\,\,\,\,\,\,\,\,\,\,\,\,\,\,\,\,\,\,\,\,\,\,\,\,\,\,\,\,\,\,\,\,\,\,\, and                             \\    
                             &                               & $i(u,t) = \left\{\begin{array}{l}1, \,\, if \,\, \langle u,v',\theta, t' \rangle \in \mathcal{A} \wedge t' \leq t \\0, \,\, otherwise \\ \end{array}\right.$                   \\												
\hline  Structural           & Active Strong Connected       & $ACC^{\theta}_{v,t} = |P(\eta_{v,t}^{\theta})|$                                                                                                                                 \\
        Diversity            & Components Count \cite{Ugander:2012} & where the function $P(V'): V' \rightarrow C$ maps the set of nodes $V'$                                                                                                  \\ 
                             &                               & to the set of strongly connected components $C$.                                                                                                                                \\ \cline{2-3}
                             & Active Strongly Connected     & $ACR^{\theta}_{v,t} = \frac{|P(\eta_{v,t}^{\theta})|}{|P(\eta_{v,t})|}$                                                                                                         \\ 
                             & Components Ratio \cite{Ugander:2012}  &                                                                                                                                                                                 \\
\hline 
\end{tabular}
\vspace{-8pt}
\label{tab:features}
\vspace{-8pt}
}
\end{table*}

Figure \ref{fig:Correlation_Gain}(c) shows the heat map for Continuous Decay of Influence (CDI). Highest gains in [1.72\% 60.34\%] are observed when $\tau_{sus} \ge 168$ and $\tau_{fos} \le 96$.

\vspace{5pt}

Figure \ref{fig:Correlation_Gain}(d) shows the heat map for Previous Reposts (PRR).  There is a slight tendency that higher values to $\tau_{fos}$ and lower values $\tau_{sus}$ result in higher when compared to the previous features, with gains in [4.76\%, 156\%].

Figure \ref{fig:Correlation_Gain}(e) presents the heat map for Closed Triads (CLT).  Higher gains in [3.33\%, 233.33\%] are spread through the majority of cells. However, we can still observe best gains distributed over the area where high values of $\tau_{sus}$ and low values of $\tau_{fos}$ are found, specially when $\tau_{sus} >= 144$ and $\tau_{fos} <= 48$.  Figure \ref{fig:Correlation_Gain}(f) shows the heat map for Clustering coefficient (CLC). Small values for both time constraints produce higher correlation gains in [4.84\%, 66.67\%].

Figure \ref{fig:Correlation_Gain}(g) presents the heat map for Hubs (HUB). We again observe the hot spots where $\tau_{sus}$ and $\tau_{fos}$ are relatively high and low respectively, with $\tau_{sus} >= 96$ and $\tau_{fos} <= 48$. Gains in [21.05\%, 488.24\%] are the highest. 

Figure \ref{fig:Correlation_Gain}(h) presents the heat map for Mutual Reposts (MUR), another cumulative measurement whose correlation increases with both time constraints. However, we can observe higher gains in [1.61\%, 35.48\%] found in the area where the values of $\tau_{sus}$ are relatively high and the values of $\tau_{fos}$ are intermediate.

Figure \ref{fig:Correlation_Gain}(i) shows the heat map for Active Strong Connected Components Count (ACC). Hot cells are found where $\tau_{sus}$ and $\tau_{fos}$ are relatively high and low respectively, mainly when $\tau_{sus} >= 168$ and $\tau_{fos} <= 24$, with gains in [9.09\%, 300\%].  Figure \ref{fig:Correlation_Gain}(j) presents the heat map for Active Strong Connected Components Ratio (ACR). The gains in [1.47\%, 45.59\%] are spread through the cells, mainly where values of $\tau_{sus}$ and $\tau_{fos}$ are both relatively small.
		
\vspace{-15pt}
\section{Classification Experiments}
\vspace{-13pt}
This section presents our classification experiments and results for the adoption prediction task, detailing training and testing sets and baselines comparisons.

\vspace{3pt}

\noindent\textbf{Training and testing}. Our 10 social influence measures are treated here as features in a machine learning task, such that we can measure their performance for adoption prediction individually and combined.  We sort our samples chronologically, using the first 90\% for training and the rest for testing (obeying causality which is neglected by some works).  We use 2 classifiers, Logistic Regression \cite{Attewell:2015} and Random Forest \cite{Attewell:2015}, but only report F1 score for Random Forest under the \textit{R60} filter, since Logistic Regression and other filters produce comparable results.

\vspace{2pt}

\noindent\textbf{Baselines}. We compare our model with 3 baselines (Influence Locality (LRC-Q) \cite{Zhang:2013}, Static Bernoulli (SB) \cite{Goyal:2010}, Complex Probability Model (CPM) \cite{Fink:2016}), to check if our method outperforms them and if the baselines improve with $\tau_{sus}$ and $\tau_{fos}$.
	
\vspace{2pt}

\noindent\textbf{Individual Feature Analysis}. Table \ref{tab:feature_performances} presents the individual classification performance of our 10 features.  As done in the previous section, we run an experiment for each combination of $\tau_{sus}$ and $\tau_{fos}$, sorting this table by the performance gain.  The time constraints boosted the performance in all cases, with gains in F1 score in [7.22\% and 23.2\%].  In the great majority of cases, $\tau_{sus}$ shows values greater than $\tau_{fos}$, repeating the correlation gain pattern.

\vspace{-7pt}		
\begin{table}[!ht]
\vspace{-7pt}		
	\centering
	\tiny{
		\caption{Baselines, Individual and Combined Feature Performances.}
		\vspace{-5pt}
		\begin{tabular}{|c|c|c|c|c|c|c|c|c|c|}
		
		  \hline \multicolumn{4} {|c|}{PNE w/ time constraints} & {PNE w/o time constraints} & \multicolumn{4} {|c|}{CLC w/ time constraints} & {CLC w/o time constraints} \\
		  \hline  $\tau_{sus}$ & $\tau_{fos}$ & F1 score   & Improv. & F1 score & $\tau_{sus}$ & $\tau_{fos}$ & F1 score & Improv. & F1 score \\ 
			\hline  168          & 24           & 0.658      & 23.2\% & 0.534    & 72           & 16           & 0.652    & 22.0\% & 0.534    \\
		
		  \hline \multicolumn{4} {|c|}{ACR w/ time constraints} & {ACR w/o time constraints} & \multicolumn{4} {|c|}{CLT w/ time constraints} & {CLT w/o time constraints} \\
			\hline  $\tau_{sus}$ & $\tau_{fos}$ & F1 score   & Impro. & F1 score & $\tau_{sus}$ & $\tau_{fos}$ & F1 score & Impro. & F1 score \\ 
			\hline  336          & 120          & 0.632      & 18.7\% & 0.532    & 144          & 8            & 0.657    & 17.3\% & 0.560    \\
		
		  \hline \multicolumn{4} {|c|}{NAN w/ time constraints} & {NAN w/o time constraints} & \multicolumn{4} {|c|}{CDI w/ time constraints} & {CDI w/o time constraints} \\
			\hline  $\tau_{sus}$ & $\tau_{fos}$ & F1 score   & Impro. & F1 score & $\tau_{sus}$ & $\tau_{fos}$ & F1 score & Impro. & F1 score \\ 
			\hline  144          & 72           & 0.689      & 15.6\% & 0.596    & 144          & 16           & 0.677    & 13.5\% & 0.596    \\

		  \hline \multicolumn{4} {|c|}{ACC w/ time constraints} & {ACC w/o time constraints} & \multicolumn{4} {|c|}{HUB w/ time constraints} & {HUB w/o time constraints} \\
			\hline  $\tau_{sus}$ & $\tau_{fos}$ & F1 score   & Impro. & F1 score & $\tau_{sus}$ & $\tau_{fos}$ & F1 score & Impro. & F1 score \\ 
			\hline  144          & 16           & 0.675      & 13.2\% & 0.596    & 120          & 16           & 0.630    & 9.99\% & 0.573    \\

		  \hline \multicolumn{4} {|c|}{PRR w/ time constraints} & {PRR w/o time constraints} & \multicolumn{4} {|c|}{MUR w/ time constraints} & {MUR w/o time constraints} \\
			\hline  $\tau_{sus}$ & $\tau_{fos}$ & F1 score   & Impro. & F1 score & $\tau_{sus}$ & $\tau_{fos}$ & F1 score & Impro. & F1 score \\ 
			\hline  72           & 8            & 0.672      & 9.82\% & 0.612    & 336          & 72           & 0.712    & 7.23\% & 0.664    \\

		  \hline \multicolumn{4} {|c|}{All w/ time constraints} & {All w/o time constraints} & \multicolumn{4} {|c|}{LRC-Q w/ time constraints} & {LRC-Q w/o time constraints} \\
			\hline  $\tau_{sus}$ & $\tau_{fos}$ & F1 score   & Impro.  & F1 score & $\tau_{sus}$ & $\tau_{fos}$ & F1 score & Impro. & F1 score \\                                      
			\hline  336          & 48           & 0.755      & 10.54\% & 0.683    & 72           & 16           & 0.657    & 8.41\% & 0.606    \\

		  \hline \multicolumn{4} {|c|}{SB w/ time constraints} & {SB w/o time constraints} & \multicolumn{4} {|c|}{CPM w/ time constraints} & {CPM w/o time constraints} \\
			\hline  $\tau_{sus}$ & $\tau_{fos}$ & F1 score   & Impro.  & F1 score & $\tau_{sus}$ & $\tau_{fos}$ & F1 score & Impro. & F1 score \\                                      
			\hline  72           & 8            & 0.675      & 8.69\%  & 0.621    & 72           & 8            & 0.689    & 12.58\% & 0.612   \\
		
			\hline 
			
		\end{tabular}
    \vspace{-9pt}		
		\label{tab:feature_performances}
		\vspace{-8pt}
	}
\end{table}
\vspace{8pt}

\noindent\textbf{Combined Feature Analysis}. In Table \ref{tab:feature_performances}, we also present the classification performance results when the 10 features are combined as ``All'', showing an improvement of 10.54\% when applying time constraints.  The observed pattern for the individual features (social influence is better measured by the measures when $\tau_{sus} > \tau_{fos}$) is found again for the features combined, with performance achieving the best improvements when $\tau_{sus} = 336$, while $\tau_{fos} = 48$.  

We interpret these results as: 1). users will start losing attention of their neighbors after 2 weeks, if they do not retweet them anymore; 2). users will no more remember the activations of their neighbors after approximately 2 days. 

\vspace{2pt}

\noindent\textbf{Performance of Baseline Methods}. Finally, Table \ref{tab:feature_performances} includes the results of baselines. The time constraints boost all performances, with gains of 8.41\% for LRC-Q, 8.69\% for SB and 12.58\% for CPM. These results highlight the effectiveness of $\tau_{sus}$ and $\tau_{fos}$, also consolidating the pattern detected before: $\tau_{sus} > \tau_{fos}$.  In addition, our model outperforms the baselines in both situations: when we use only an individual feature such as MUR, and when we use all features combined, with improvements in [3.33\%, to 9.6\%] (compared with CPM).

\vspace{-10pt}
\section{Conclusion}
\vspace{-9pt}
In this paper, we introduce a pair of time constraints to show how the dynamic graphs produced by them better capture the influence between users over time (specially when $\tau_{sus}$ and $\tau_{fos}$ are relatively high and low respectively).  We validate our model under diverse conditions, detailing how it outperforms the state-of-the-art methods that aim to predict users' adoption. We also demonstrate how these constraints can be used to improve the performance of other approaches, enabling practical usage of the concepts for social influence prediction.  

\vspace{-13pt}
\section*{Acknowledgments}
\vspace{-8pt} 
\scriptsize
Some of the authors of this paper are supported by CNPq-Brazil, AFOSR Young Investigator Program
(YIP) grant FA9550-15-1-0159, ARO grant W911NF-15-1-0282, and the DoD Minerva program.

\end{document}